\documentstyle[aps,preprint]{revtex}

\widetext
\draft
\tighten
\def\abstract{\if@twocolumn
\section*{Resumen}
\else \small
\begin{center}
{\bf Resumen\vspace{-.5em}\vspace{0pt}}
\end{center}
\quotation
\fi}

\def\thebibliography#1{\section*{Referencias\markboth
{REFERENCIAS}{REFERENCIAS}}
\list
{[\arabic{enumi}]}{\settowidth\labelwidth{[#1]}
\leftmargin\labelwidth
\advance\leftmargin\labelsep
\usecounter{enumi}}
\def\newblock{\hskip .11em plus .33em minus -.07em}
\sloppy
\sfcode`\.=1000\relax}

\begin{document}

\preprint{EFUAZ FT-96-27}

\title{Las Construcciones de Dirac y de Majorana
(Agenda para Estudiantes)\thanks{Enviado a ``Investigaci\'on
Cientifica".}}

\author{{\bf Valeri V. Dvoeglazov}}

\address{
Escuela de F\'{\i}sica, Universidad Aut\'onoma de Zacatecas \\
Antonio Doval\'{\i} Jaime\, s/n, Zacatecas 98068, ZAC., M\'exico\\
Correo electronico:  VALERI@CANTERA.REDUAZ.MX}

\date{5 de julio de 1996}

\maketitle

\abstract{El presente trabajo es una breve revisi\'on del desarollo
reciente de la mec\'anica cu\'antica relativista en el espacio de
representaci\'on $(1/2,0)\oplus (0,1/2)$.  Est\'a dirigida a
estudiantes de licenciatura, pero tambi\'en puede ser \'util a
estudiantes de posgrado en f\'{\i}sica de part\'{\i}culas y
teor\'{\i}a cu\'antica de campos.

\bigskip

The present work is a brief review of the recent development of the
relativistic quantum  mechanics in the $(1/2)\oplus (0,1/2)$
representation space.  It is  directed for  B. Sc. students  of
Mexican Universities, but can be also useful for students graduated
in  particle physics and quantum field theory.}


\newpage

\setlength{\baselineskip}{24pt}

La construcci\'on de Dirac~\cite{Dirac0} para fermiones es
conocida por todos los f\'{\i}sicos. Es muy \'util porque nos permite
describir  fen\'omenos con part\'{\i}culas cargadas, las que en nuestra
disposici\'on en los aceleradores. Se ha dado
considerable atenci\'on  desde los
primeros principios  a esta construcci\'on en
art\'{\i}culos precedentes para los estudiantes~\cite{DV-IC1,DV-IC2}. En
los \'ultimos sesenta a\~nos los f\'{\i}sicos prestaron poca
atenci\'on a la construcci\'on de Majorana~\cite{Majorana} para
part\'{\i}culas neutras tales como neutrino y fot\'on; existen tres
art\'{\i}culos~\cite{MLC,Tokuoka} de cierta antig\"uedad que hacen
referencia a estos problemas.  Adem\'as, los modelos diferentes del modelo
del electr\'on de Dirac han sido propuestos por Pauli, Weisskopf y
Markov~\cite{Pauli,Markov1,Markov2}.  Recientemente, gracias a muchas
discrepancias entre experimento y el modelo est\'andar de Weinberg, Salam
y Glashow en f\'{\i}sica del neutrino, ademas a unas causalidades,
rigurosa consideraci\'on del espacio de representaci\'on $(j,0)\oplus
(0,j)$ basada en la teor\'{\i}a del grupo de Lorentz ha sido empezada por
D.  V.  Ahluwalia~\cite{DVA1,DVA2,DVA3}, G.  Ziino, A.  O.
Barut~\cite{Ziino}, y por V. V. Dvoeglazov~\cite{DVO1,DVO11,DVO2,DVO3}.

Continuando quisiera aclarar, si el
lector quiere entender nuestras ideas, \'el tiene que acostumbrarse a
pensar que la construcci\'on de Dirac es la verdad, s\'olo la verdad
pero {\bf no toda la verdad}.

Los postulados que usamos son los siguientes:

\begin{itemize}

\item
La relaci\'on dispersional relativista $E^2/c^2 - {\bf p}^2 =m^2 c^2$ es
valida para los estados de part\'{\i}culas observables; $c$ es la
velocidad de la luz.

\item
Los 2-espinores derechos e izquierdos en el espacio $(j,0)\oplus (0,j)$,
donde $j$ es el esp\'{\i}n de part\'{\i}cula, se transforman de acuerdo
con reglas de Wigner~\cite{Wigner1,Wigner2}; $\bbox{\varphi}$ son los
par\'ametros del {\tt boost} dado,
\begin{mathletters} \begin{eqnarray}
(j,0):&&\quad\phi_{_R} (p^\mu)\, = \,\Lambda_{_R} (p^\mu \leftarrow
\overcirc{p}^\mu)\,\phi_{_R} (\overcirc{p}^\mu) \, = \, \exp (+\,{\bf J}
\cdot {\bbox \varphi}) \,\phi_{_R} (\overcirc{p}^\mu)\quad,\\
(0,j):&&\quad \phi_{_L} (p^\mu)\, =\, \Lambda_{_L} (p^\mu \leftarrow
\overcirc{p}^\mu)\,\phi_{_L} (\overcirc{p}^\mu) \, = \, \exp (-\,{\bf J}
\cdot {\bbox \varphi})\,\phi_{_L} (\overcirc{p}^\mu)\quad,\label{boost0}
\end{eqnarray}
\end{mathletters}
que son relacionadas con el postulado (1), vease Ryder~\cite{Ryder}.

\item
En el sistema de referencia donde el momento lineal de la part\'{\i}cula
es igual a cero tenemos la relaci\'on
de Ryder-Burgard~\cite{Ryder,DVA1,DVA3,DVO1,DVO11,DVO3} entre los espinores
izquierdos y derechos.  Esta relaci\'on se basa en la
declaraci\'on~\cite{Ryder}: ``{\it When a particle is at rest, one
cannot define its spin as either left- or right-handed, so $\phi_{_R}
(0) = \phi_{_L} (0)$}. Vease mas adelante para la relaci\'on generalizada
y su discusi\'on.

\end{itemize}

Como lo notific\'o D. V. Ahluwalia~\cite{DVA3} aparte del {\tt bispinor} de
Dirac $\Psi (p^\mu) = column (\phi_{_R} (p^\mu),\quad \phi_{_L} (p^\mu))$
es posible considerar los bispinores auto/contr-auto conjugados desde el
principio porque sabemos~\cite{Ryder} que $\Theta_{[j]} \phi_{_L}^\ast$
se transforma como los espinores derechos y $\Theta_{[j]} \phi_{_R}^\ast$,
como los espinores izquierdos, $\Theta_{[j]}$ es el operador de
reversi\'on de tiempo para el esp\'{\i}n $j$. Entonces, podemos definir
cuadriespinores
\begin{equation} \lambda(p^\mu)\,\equiv \pmatrix{ \left (
\zeta_\lambda\,\Theta_{[j]}\right )\,\phi^\ast_{_L}(p^\mu)\cr
\phi_{_L}(p^\mu)} \,\,,\quad \rho(p^\mu)\,\equiv \pmatrix{
\phi_{_R}(p^\mu)\cr
\left ( \zeta_\rho\,\Theta_{[j]}\right )^\ast
\,\phi^\ast_{_R}(p^\mu)} \,\,\quad ,\label{sp-dva}
\end{equation}
donde $\zeta_\lambda$ y $\zeta_\rho$ son los factores de fase
que se fijan por las condiciones de auto/contr-auto conjugaci\'on de carga
el\'ectrica (o, sus generalizaciones para los esp\'{\i}nes
enteros~\cite[ec.(12)]{DVA3})
\begin{equation} S^c_{[j]} \lambda^{S,A} (p^\mu) = \pm
\lambda^{S,A} (p^\mu)\,.
\end{equation}
Los bispinores tienen
propiedades extra\~nas: a) no son funciones propias del operador
de helicidad en el espacio $(j,0)\oplus (0,j)$; b) no son funciones
propias del operador de paridad, que en la representaci\'on de Weyl tiene
la forma $$S^s_{[j]} = e^{i\vartheta^s_{[j]}} \pmatrix{0&\openone_j\cr
\openone_j &0\cr};$$ c) los estados que describen son estados {\tt
bi-ortonormales} en sentido matem\'atico; d) los estados {\bf no} son
estados propios del operador hamiltoniano de Dirac.  Los bispinores
pueden satisfacer dos tipos de ecuaciones.  Las primeras han sido
dadas por D. V.  Ahluwalia para espinores
$\lambda^{S,A}$~\cite{DVA3}
\begin{eqnarray} \pmatrix{-\,\openone &
\zeta_\lambda\,\exp\left( {\bf J}\,\cdot \bbox{\varphi}\right )
\,\Theta_{[j]}\,{\mit\Xi}_{[j]}\, \exp\left( {\bf J}\,\cdot \bbox{\varphi}
\right )\cr \zeta_\lambda\,\exp\left(-\, {\bf
J}\,\cdot\bbox{\varphi}\right) \,{\mit\Xi}^{-1}_{[j]}\,\Theta_{[j]}\,
\exp\left(- \,{\bf J}\,\cdot\bbox{\varphi} \right) & -\,\openone}\,\lambda
(p^\mu)\,=\,0\,,\label{genweq1} \end{eqnarray}
y por V. V. Dvoeglazov para espinores $\rho^{S,A}$~\cite{DVO3}
\begin{eqnarray}
\pmatrix{-\,\openone & \zeta_\rho^*\,\exp\left(
{\bf J}\,\cdot \bbox{\varphi}\right )\,
{\mit\Xi}_{[j]}^{-1}\,\Theta_{[j]}\, \exp\left( {\bf J}\,\cdot
\bbox{\varphi} \right )\cr
\zeta_\rho^*\,\exp\left(-\, {\bf
J}\,\cdot\bbox{\varphi}\right) \,\Theta_{[j]}\,{\mit\Xi}_{[j]}\,
\exp\left(- \,{\bf J}\,\cdot\bbox{\varphi}
\right) & -\,\openone}\,\rho (p^\mu)\,=\,0\,.\label{genweq2}
\end{eqnarray}
La deducci\'on de las ecuaciones ha sido presentada
en ref.~\cite{DVA3,DV-IC1} y est\'a basada en la relaci\'on de
Ryder-Burgard de la siguiente forma:  \begin{equation} \left
[\phi_{_{L,R}}^h (\overcirc{p}^\mu)\right ]^* = {\mit\Xi}_{[j]}
\phi_{_{L,R}}^h (\overcirc{p}^\mu)\quad, \end{equation}
donde $\Xi_{[j]}$
es la matriz que conecta el bispinor de momento cero con su complejo
conjugado. Esta ecuaci\'on no se puede representar en la forma covariante
$[\Gamma^{\mu\nu} p_\mu p_\nu +m\Gamma^\mu p_\mu -2m^2\openone ]\lambda
=0$ porque, como mencion\'o Ahluwalia, ``it turns out that
$\Gamma^{\mu\nu}$ and $\Gamma^\mu$ do not transfrom as Poincar\`e
tensors".  Adem\'as, despues de aplicar el sistema para calculos
analiticos en computadora MATHEMATICA 2.2 se puede convencer que aparte
de las soluciones que satisfacen el postulado (1), existen soluciones de
la ecuaci\'on para $j=1/2$~\cite[ec.  (31)]{DVA3} con la relaci\'on
dispersional $p_0 = -2m \pm \sqrt{{\bf p}^{\,2} +m^2}$, a las cuales
tenemos que proponer la interpretaci\'on.  El otro camino para encontrar
las ecuaciones de los bispinores auto/contr-auto conjugados es el que usa
la forma de la relaci\'on de Ryder-Burgard siguiente:  \begin{equation}
\left [\phi_L^h (\overcirc{p}^\mu)\right ]^* = (-1)^{1/2-h} e^{-i(\theta_1
+\theta_2)} \Theta_{[1/2]} \phi_L^{-h} (\overcirc{p}^\mu)\quad,
\end{equation}
que relaciona los espinores de helicidad opuesta.  Obtenemos
el conjunto de ecuaciones covariantes:
\begin{mathletters} \begin{eqnarray} i
\gamma^\mu \partial_\mu \lambda^S (x) - m \rho^A (x) &=& 0 \quad,
\label{11}\\
i \gamma^\mu \partial_\mu \rho^A (x) - m \lambda^S (x) &=& 0 \quad,
\label{12}
\end{eqnarray}
\end{mathletters}
para los estados con la energ\'{\i}a positiva, y
\begin{mathletters}
\begin{eqnarray}
i \gamma^\mu \partial_\mu \lambda^A (x) + m \rho^S (x) &=& 0\quad,\\
\label{13}
i \gamma^\mu \partial_\mu \rho^S (x) + m \lambda^A (x) &=& 0\quad,
\label{14}
\end{eqnarray}
\end{mathletters}
para los estados con energ\'{\i}a negativa. Estas ecuaciones se pueden
escribir en la forma $8$- dimensional y ellas tienen las relaciones
dispercionales $p_0 = \pm \sqrt{{\bf p}^{\,2} +m^2}$ \'unicamente.
La interpetaci\'on f\'{\i}sica de las ecuaciones (\ref{11}-\ref{14})
es parecida a la interpretaci\'on de los art\'{\i}culos de
Barut y Ziino~\cite{Ziino}.  Por ejemplo, el pricipal resultado
es que los estados auto/contr-auto conjugados de carga
el\'ectrica aunque son neutras en el sentido de las interacciones
electromagn\'eticas no son neutras con respecto de carga quiral, el
nuevo tipo del operador de carga que anticomuta con el operador de carga
el\'ectrica. El primero quien lo predijo hace mucho tiempo fue R. E.
Marshak, y el operador de carga quiral ha sido conectado con las
transformaciones de norma $\cos\alpha \pm i\gamma^5 \sin\alpha$.
Analizando la forma de los bispinores (\ref{sp-dva}) podemos deducir tres
tipos m\'as de interacci\'on de norma.  Quisiera mencionar que las
ecuaci\'ones del tipo (\ref{11}-\ref{14}) fueron presentadas por primera
vez por M. Markov~\cite{Markov1,Markov2}, el autor de conocida idea de
friedmones.  El fue tambi\'en el primero qui\'en expres\'o la idea que el
problema de jerarqu\'{\i}a (del espectro de leptones, en otras palabras)
tiene que ser resuelta en base al an\'alisis de las soluciones de ``la
ecuaci\'on de Dirac" con ocho componentes y sus interacciones con el
vector potencial (v\'ease tamb\'en Barut~\cite{Barut}).  La conexi\'on de
la ecuaci\'on de Dirac con ocho componentes con la construcci\'on de
Majorana ha sido estudiada por Tokuoka~\cite{Tokuoka}.

Finalmente, queremos recordar que diferentes tipos de las teor\'{\i}as
invariantes con respecto al grupo de Lorentz extendido (que toman en
cuenta las operaciones de simetr\'{\i}as  discretas), las
teor\'{\i}as que recibieron el nombre del tipo de Bargmann, Wightman y
Wigner~\cite{DVA1},  han sido consideradas
por varios autores~\cite{Wigner2,Nigam,Gelf,Sengup,Fush,San}.
Nuestras construcciones pertencen a este tipo.
Adem\'as, como se indicaron~\cite{Sengup,Fush,Simon,San} en
aplicaci\'on a neutrino, aunque la part\'{\i}cula puede tener masa cero
(de la ecuaci\'on de Klein-Gordon), sin embargo  la ecuaci\'on de
primer orden parecida a la ecuaci\'on de Dirac puede contener el t\'ermino
con alg\'un par\'ametro (`masa variable' como lo llamo el profesor
Fushchich) de la dimensi\'on de masa.
Por ejemplo, \begin{equation} \left
[i\gamma^\mu \partial_\mu +m (1 \pm \gamma_5) \right ] \psi (x^\mu) =
0\quad.  \end{equation} Otros tipos de ecuaciones no tienen la forma
covariante explicita.

En base a los trabajos que mencion\'e en esta peque\~na revisi\'on
podemos delinear nuestros planes en el futuro pr\'oximo:

\begin{itemize}

\item
Por consideraci\'on a las transformaciones de norma m\'as generales para
los espinores $\lambda$ y $\rho$  deducir el modelo de Weinberg, Salam y
Glashow.

\item
En base de las ideas de Dowker~\cite{Dowker}, Evans y Vigier~\cite{Evans}
entender qu\'e es esp\'{\i}n.

\item
Deducir el espectro de leptones en base de la desarollo matem\'atico de
las ideas de Markov~\cite{Markov2} y Barut~\cite{Barut}.

\item
Investigar la estructura matem\'atica del {\tt espacio de Fock} en base a
las ideas de Dirac~\cite{Dirac}.

\item
En base de la consideraci\'on de los espinores en el sistema de referencia
con el momento lineal cero entender que es {\tt color}, el n\'umero
cu\'antico de los {\tt quarks} y {\tt gluones} en cromodin\'amica
cu\'antica y por qu\'e los estados `de color'  no pueden ser observables
directamente.  Deducir la teor\'{\i}a de norma basada en el
grupo $SU_c (3)$, la cromodin\'amica cu\'antica.

\item
Incluir la gravitaci\'on para unificar todas las interacci\'ones.

\end{itemize}

\smallskip

Como conclusi\'on de mis tres art\'{\i}culos en
``Investigaci\'on Cient\'{\i}fica" sobre la mec\'anica cu\'antica
relativista: aunque el modelo est\'andar de Glashow, Weinberg y Salam
y otros modelos de norma fueron muy utiles en predicciones
fenomenologicas, ellos todav\'{\i}a no pueden ser considerados
como base para una teor\'{\i}a completa. Los campos cu\'anticos
con  $2(2j+1)$ componentes construidos en base de los postulados
de teor\'{\i}a de relatividad y el principio de causalidad contienen
ciertas simetr\'{\i}as que junto con los postulados de los modelos de
norma pueden ser la base para la construcci\'on de
una teor\'{\i}a unificada.


\bigskip

Agradezco mucho a  los doctores Dharam V. Ahluwalia y Anatoly F. Pashkov
por su apoyo y paciente introducci\'on  en estas materias.
Agradezco el apoyo en la ortograf\'{\i}a espa\~nola
del Sr.  Jes\'us Alberto C\'azares.

\end{document}